\documentclass[letterpaper, 10 pt, conference]{icnrconf}
\pdfoutput=1
\usepackage{graphicx}
\usepackage[T1]{fontenc}
\usepackage[utf8]{inputenc}
\usepackage[UKenglish]{babel}

\usepackage{float} 
\usepackage{multirow}
\usepackage{booktabs}
\usepackage{lscape}
\usepackage{amssymb}
\usepackage{amsmath}
\usepackage[]{algorithm2e}

\makeatletter
\let\MYcaption\@makecaption
\makeatother
\usepackage[font=footnotesize]{subcaption}
\makeatletter
\let\@makecaption\MYcaption
\makeatother

\usepackage{pgfplots} 
\usepackage{adjustbox}

\makeatletter 
\let\NAT@parse\undefined
\makeatother
\usepackage[linktoc=page]{hyperref}
\usepackage[nameinlink]{cleveref}
\usepackage{cite}
\newcounter{inlineenum}
\newenvironment{inlineenum}
{\unskip\ignorespaces\setcounter{inlineenum}{0}%
	\renewcommand{\item}{\refstepcounter{inlineenum}{\textit{\theinlineenum})~}}}
{\ignorespacesafterend}

\Crefname{equation}{Eq.}{Eqs.}
\Crefname{figure}{Fig.}{Figs.}
\Crefname{tabular}{Tab.}{Tabs.}

\hypersetup{
	pdfauthor = {Tim Sziburis, Markus Nowak, Davide Brunelli},
	pdftitle = {kNN Learning Techniques for Proportional Myocontrol in Prosthetics},
	pdfsubject = {kNN Learning Techniques for Proportional Myocontrol in Prosthetics}
}

\hypersetup{pdfborder = {0 0 0}}

\IEEEoverridecommandlockouts                             
\title{\LARGE \bf kNN Learning Techniques for Proportional Myocontrol in Prosthetics}

\author{Tim Sziburis$^1$, Markus Nowak$^2$, Davide Brunelli$^3$
\thanks{$^1$ T. Sziburis is with the High Precision Alignment Technologies Section, CERN, 1211 Geneva 23, Switzerland, and was with the German Aerospace Center (DLR), Robotics and Mechatronics Center (RMC), M\"unchner Str. 20, 82234 We\ss ling, Germany ({\tt\small tim.sziburis@cern.ch})}
\thanks{$^2$ M. Nowak is with the German Aerospace Center (DLR), Robotics and Mechatronics Center (RMC), M\"unchner Str. 20, 82234 We\ss ling, Germany}
\thanks{$^3$ D. Brunelli is with the Department of Industrial Engineering, DII, University of Trento, 38123 Trento, Italy}
}

\begin{document}
\bstctlcite{IEEEexample:BSTcontrol}
\maketitle

\begin{abstract}
This work has been conducted in the context of pattern-recognition-based control for electromyographic prostheses. It presents a k-nearest neighbour (kNN) classification technique for gesture recognition, extended by a proportionality scheme.
The methods proposed are practically implemented and validated. Datasets are captured by means of a state-of-the-art 8-channel electromyography (EMG) armband positioned on the forearm. Based on this data, the influence of kNN's parameters is analyzed in pilot experiments.
Moreover, the effect of proportionality scaling and rest thresholding schemes is investigated.
A randomized, double-blind user study is conducted to compare the implemented method with the state-of-research algorithm Ridge Regression with Random Fourier Features (RR-RFF) for different levels of gesture exertion. The results from these experiments show a statistically significant improvement in favour of the kNN-based algorithm.
\end{abstract}

\section{Motivation and Related Work}
The kNN learning scheme has been applied for myoelectric control of prosthetic devices several times \cite{5704586, 7591042, Geethanjali2015, 6487520}. So far, kNN was utilized for sole classification as an intention detection method based on EMG signals.
In preliminary experiments, kNN showed promising results in gesture detection referring to success rate (SR) as well as generalizability. It is considered as robust (i.~a., against electrode shift \cite{7860925} and sampling frequency variation \cite{8122765}). Further benefits are the algorithm's incrementality and low level of implementation complexity. As kNN is an instance-based machine learning algorithm, training of an explicit model is not necessary.

Since proportionality is a key feature in myocontrol, regression algorithms gain more popularity \cite{Ameri, Gijsberts}; moreover several attempts were made to combine classification concepts with proportional scaling in EMG-based intention detection: LDA \cite{Lobov, Hudgins}, neural networks \cite{Scheme, Simon} and common spatial patterns \cite{Amsuess2015} have been used to control the velocity based on the signal intensity.
To the authors' knowledge, kNN was not adapted as a proportional scheme so far. In this study, we developed such a scheme investigating several modalities to include proportionality.

\section{Conceptual Approach}
It is assumed that the intensity of an exerted gesture is approximately proportional to the amplitude of the signal (mean of all channels of the 8-channel EMG signal). 
We intend to use this property for proportional scaling of gestures classified by kNN.
The rectified sensor reading is divided into a normalized signal and the signal strength (normalization factor), while the former is used for gesture prediction, and the latter for scaling under the assumption of linear correlation of signal strength and proportional intent for the particular gesture.
The rest gesture is treated independently. The mean magnitude of the rest samples gathered during training ($t_0$) is taken as baseline for rest. If the threshold $t=g\cdot t_0$ (gain $g$) is exceeded, a gesture is not classified as rest anymore. An increase in $t$ reduces unwanted activations.

However, a higher level of $t$ requires the user to exert higher forces to activate a gesture and therefore also leads to a lower resolution of proportionality. There is a trade-off between suppressing unwanted activations and providing a high level of resolution (maintaining the maximum).
We therefore introduce a divisor $d$ which scales the proportionality function offset $m_0=\frac{t}{d}$; but not the threshold for rest $t$ itself. Different configurations are tested in pilot tests.

\section{Control Method Adaptation}
The training process comprises:
\begin{inlineenum}
	\item capturing training data,
	\item calculating class magnitude means for proportionality scaling,
	\item normalization of this data,
	\item block-wise cross-validation for obtaining the optimal $k$ in terms of accuracy.
\end{inlineenum}

The prediction process is structured as follows:
\begin{inlineenum}
	\item rest thresholding,
	\item normalization of new sample,
	\item calculating $k$ nearest neighbours of the new sample,
	\item applying distance weighting on the $k$ selected neighbouring samples,
	\item executing kNN classification by majority voting, and
	\item signal magnitude analysis and scaling the prediction proportionally.
\end{inlineenum}

For the preparation of the user study, we conducted a two-step method adaptation (analyzing seven actions: rest, power grasp, point, wrist flexion/extension, pronation/supination). First, we performed offline cross-validation to determine the most suited kNN parameters ($k$, distance metric and weighting factor). This was followed by a single-subject pilot to evaluate different ways of introducing proportionality.

\subsection{Cross-Validation}
The accuracy of kNN with varying parameter sets in block-wise cross-validation was determined in an offline study on different datasets from a single subject. In the case of low $k$ ($k$ relative to the total numbers of training samples until 5--10\%), neither the metric nor the weighting was of high importance as long as applying a Minkowski-based distance, yielding 98--100\% correct classifications. Results were consistently worse with Mahalanobis distance. For higher $k$s the Euclidean norm turned out to be the best choice, together with a weighting of $\frac{1}{d^2}$. This configuration is chosen in the subsequent sections, with $k=1$.

\subsection{Pilot Tests on One Subject}
Rest thresholding (introduction of $t$) increased the SR for non-overlapping classes, as misclassifications with rest could be reduced. Maxima were achieved for $g=2.5$. 

For by trend overlapping classes applying a  divisor $d$ enabled low-intensity gestures to be more easily exerted and increased the performance.
Together with normalization, higher $d$s guaranteed that the necessary force effort to exert a full gesture does not noticeably exceed the particular training magnitudes.
The originally discovered problem that with a higher $d$ low-intensity gestures got less reachable was effaced with normalization.
Averaged over all datasets, the maximum SR $\left( 93\%\pm6\% \right)$ was achieved for $d=5$.

\subsection{User Study}
A user study with 12 subjects was conducted for evaluating the suitability in practical scenarios (using $t=2.5\cdot t_0$ and $d=5$). After signing consent forms, the participants sat in front of a screen in a standardized pose. For the training process they followed a visual stimulus, performing a repetitive series of actions, comprising the four gestures power grasp, point, wrist flexion/extension. In the prediction/test phase they were asked to follow the stimulus again, viewing a visual feedback. The goals were not only at full activation (as during training), but at levels of $33\%$ and $67\%$ as well. They were also randomized between subjects. We compared the performance in terms of SR for kNN with a regression algorithm, namely RR-RFF \cite{Gijsberts}, followed up by a statistical analysis using a one-way ANOVA on SR.

\section{Results}

\begin{figure}[h]
	\centering
	\begin{subfigure}{0.31\linewidth}
		\centering
		\includegraphics[height=4.4cm]{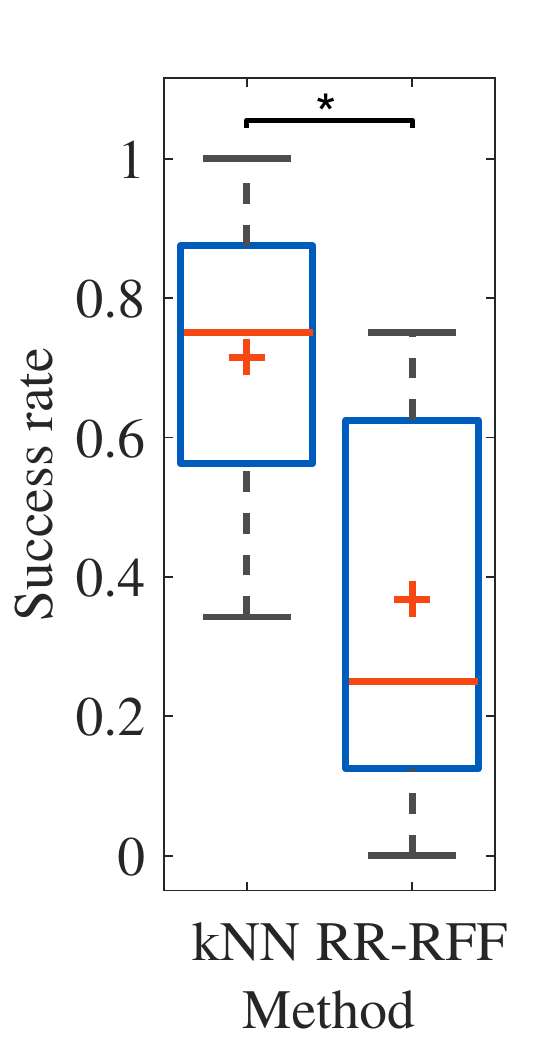}
		\caption{Overall significance}
		\label{fig:study:r2}
	\end{subfigure}
	\begin{subfigure}{0.67\linewidth}
		\centering
		\includegraphics[height=4.4cm]{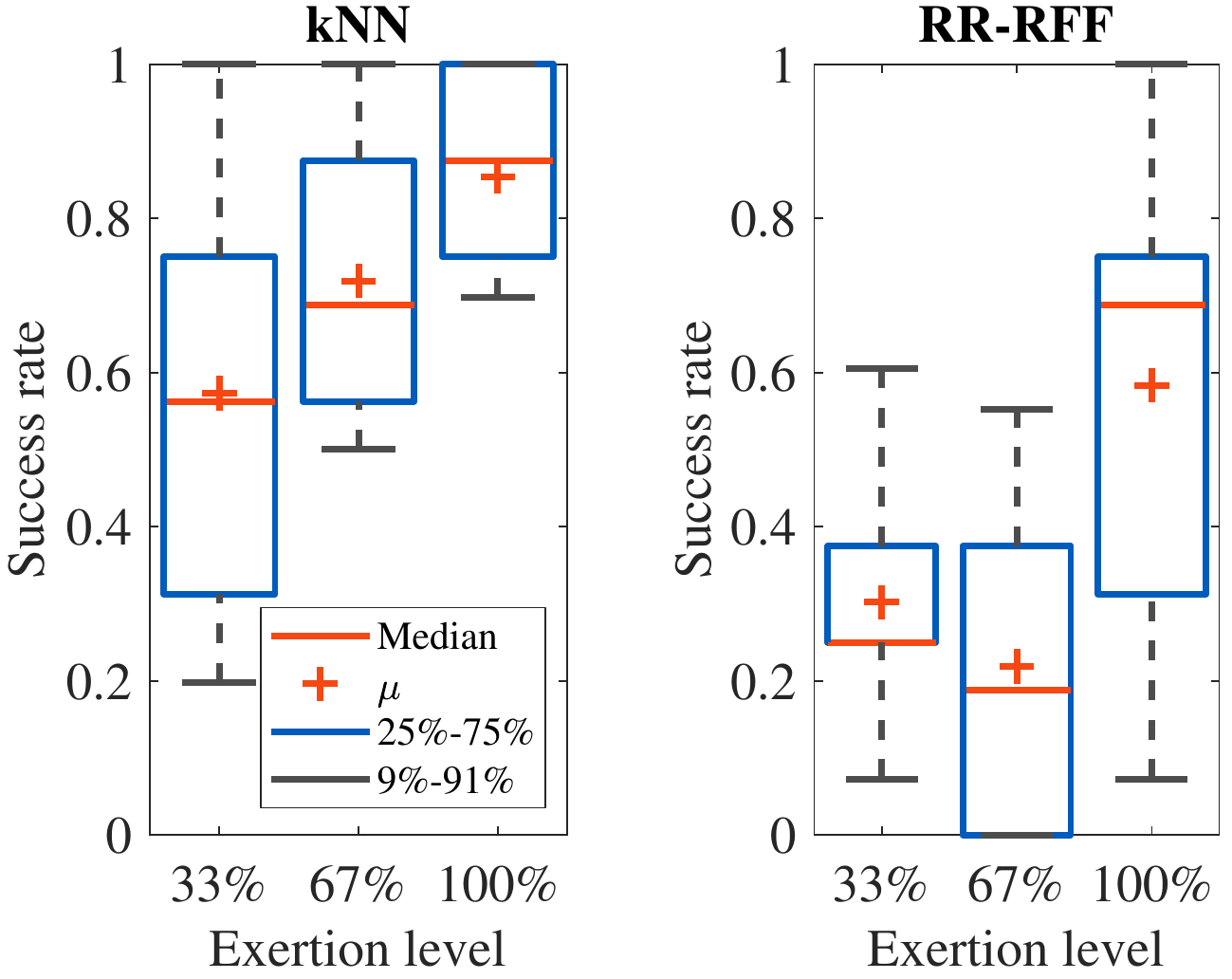}
		\caption{Dependence on exertion level}
		\label{fig:study:m2}
	\end{subfigure}\\
	\caption{User study SR results, comparing kNN and RR-RFF.}%
	\label{fig:study}%
\end{figure}

ANOVA (level of significance 5\%) reveals a stochastic significant difference in favor of kNN compared to RR-RFF, see \Cref{fig:study:r2}. The achieved SRs for different intensity levels of gesture exertion in \Cref{fig:study:m2} shows no level where RR-RFF would have outperformed kNN in median or mean of SR.

RR-RFF seems to perform well only on the intensity level it was trained with. At lower levels the performance drops drastically. This drop is less severe for kNN, where it is potentially due to training on full intensity level, and a high rest threshold causing movements with low signal amplitudes being classified as rest. The effect might be curtailed by user-specific parameter adjustments, and a learning process with subjects getting used to the algorithm's specific behaviour.

\section{Conclusions and Outlook}
This paper evaluated the extension of kNN classification by proportionality scaling for intent detection by using a state-of-the-art 8-channel myocontrol armband.
kNN can be a means towards better generalization capabilities to improve user satisfaction in prosthetics. In comparison to state-of-the-art RR-RFF, the algorithm does not involve complex numeric operations, particularly during training, and has few parameters to be tuned. An appropriate parameterization was validated in a user study.
Besides higher SR, the overall result of kNN showed a lower standard deviation than in RR-RFF, which leads to the assumption that kNN performs more stable and robust with less nondeterminism in the algorithm's behaviour. Other classification techniques may also benefit from the presented proportionality scheme. In contrast to kNN, RR-RFF allows the simultaneous detection of mixed gestures, though.
The low complexity of kNN can be of high importance for embedded system implementations. In this context, special attention must be paid to the prediction phase of standard kNN where each sample has to be related to each single other sample. This will be treated in further work.


\bibliographystyle{IEEEtranICNR}
\bibliography{ms}

\end{document}